\documentclass[a4paper,twoside,twocolumn,aps,prl,superscriptaddress,showpacs]{revtex4}
\usepackage{graphicx}
\usepackage{amsmath}
\usepackage{xcolor}
\newcommand{\figref}[1]{Fig.~\ref{#1}}
\begin{document}
\title{Quantum-to-Classical Transition in Cavity Quantum Electrodynamics}
\author{J.~M.~Fink}
\affiliation{Department of Physics, ETH Z\"urich, CH-8093, Z\"urich, Switzerland.}
\author{L.~Steffen}
\affiliation{Department of Physics, ETH Z\"urich, CH-8093, Z\"urich, Switzerland.}
\author{P.~Studer}
\affiliation{Department of Physics, ETH Z\"urich, CH-8093, Z\"urich, Switzerland.}
\author{Lev S.~Bishop}
\affiliation{Department of Physics, Yale University, New Haven, CT 06520, USA.}
\author{M.~Baur}
\affiliation{Department of Physics, ETH Z\"urich, CH-8093, Z\"urich, Switzerland.}
\author{R.~Bianchetti}
\affiliation{Department of Physics, ETH Z\"urich, CH-8093, Z\"urich, Switzerland.}
\author{D.~Bozyigit}
\affiliation{Department of Physics, ETH Z\"urich, CH-8093, Z\"urich, Switzerland.}
\author{C.~Lang}
\affiliation{Department of Physics, ETH Z\"urich, CH-8093, Z\"urich, Switzerland.}
\author{S.~Filipp}
\affiliation{Department of Physics, ETH Z\"urich, CH-8093, Z\"urich, Switzerland.}
\author{P.~J.~Leek}
\affiliation{Department of Physics, ETH Z\"urich, CH-8093, Z\"urich, Switzerland.}
\author{A.~Wallraff}
\affiliation{Department of Physics, ETH Z\"urich, CH-8093, Z\"urich, Switzerland.}
\date{\today}
\begin{abstract}
The quantum properties of electromagnetic, mechanical or other harmonic oscillators can be revealed by investigating their strong coherent coupling to a single quantum two level system in an approach known as cavity quantum electrodynamics (QED). At temperatures much lower than the characteristic energy level spacing the observation of vacuum Rabi oscillations or mode splittings with one or a few quanta asserts the quantum nature of the oscillator. Here, we study how the classical response of a cavity QED system emerges from the quantum one when its thermal occupation -- or effective temperature -- is raised gradually over 5 orders of magnitude. In this way we explore in detail the continuous quantum-to-classical crossover and demonstrate how to extract effective cavity field temperatures from both spectroscopic and time-resolved vacuum Rabi measurements.\end{abstract}
\pacs{42.50.Pq, 03.67.Lx, 07.20.Dt, 44.40.+a} \maketitle

Cavity QED \cite{Haroche2007} enables the study of the nature of matter light interactions in exquisite detail. It realizes an open quantum system in which the coupling to the environment is highly controllable. In a circuit realization of cavity QED \cite{Wallraff2004b}, we carefully investigate the quantum-to-classical transition of a harmonic oscillator strongly coupled to a two level system by increasing the effective oscillator temperature. From measured vacuum Rabi splitting spectra and from time-resolved vacuum Rabi oscillations we consistently extract effective cavity field temperatures between 100~\textrm{mK} and a few \textrm{K} using a quantum master equation model as suggested in Ref.~\cite{Rau2004a}.  The dissipative quantum-to-classical crossover of a field mode coupled to a qubit was also studied theoretically in Ref.~\cite{Everitt2009a}. The emergence of classical physics from quantum mechanics and the role of decoherence in this process is an important subject of current research \cite{Schlosshauer2007}.

\begin{figure}[b!]
\includegraphics[width=0.9 \columnwidth]{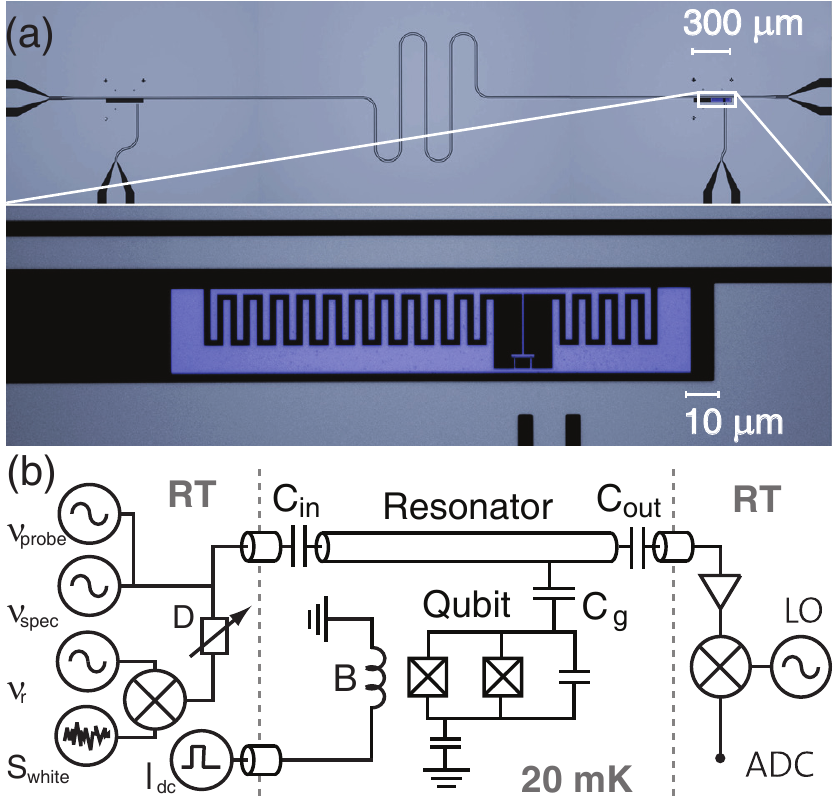}
\caption{(a), Coplanar microwave resonator with two qubit slots and flux bias lines (top) and a single embedded qubit (violet) shown on an enlarged scale (bottom). (b), Circuit diagram of setup. Input (left) is at room temperature (RT) with microwave sources for measurement ($\nu_\textrm{probe}$), qubit spectroscopy and qubit drive ($\nu_{\textrm{spec}}$), variable attenuation ($D$) quasi thermal field source ($S_\textrm{white}$) at $\nu_\textrm{r}$ and an arbitrary waveform generator for fast flux biasing ($I_\textrm{dc}$) through the on-chip flux line ($B$). At 20 \textrm{mK} the qubit is coupled via a capacitance $C_\textrm{g}$ to the transmission line resonator between the capacitances $C_\textrm{in}$ and $C_\textrm{out}$. The transmitted microwave tone is amplified, down-converted with a local oscillator (LO) and digitized (ADC).} \label{fig1}
\end{figure}

In our experiments a superconducting artificial atom is strongly coupled to a coplanar transmission line resonator to realize a circuit QED system \cite{Wallraff2004b}.  
The transmon qubit \cite{Koch2007} consists of two superconducting aluminum thin films weakly connected by two Josephson tunnel junctions. Its ground $|g\rangle$ and excited state $|e\rangle$ differ by the specific charge configuration of the two superconducting islands. The transition frequency $\nu_\textrm{g,e}\simeq\sqrt{8 E_\textrm{C} E_\textrm{J}(\Phi)}-E_\textrm{C}$ is determined spectroscopically. Here, $E_\textrm{C}/h \approx0.502 \, ~\textrm{GHz}$ is the single electron charging energy, $E_\textrm{J}(\Phi)=E_{\rm{J,max}}|\cos{(\pi \Phi/\Phi_{0})}|$ the flux controlled Josephson energy with $E_{\textrm{J,max}}/h \approx 14.4 \, \textrm{GHz}$ and $\Phi_0$ is the superconducting flux quantum. The cavity has a bare resonance frequency of $\nu_\textrm{r}\approx 6.44 \, \textrm{GHz}$ and a coupling limited photon decay rate of $\kappa/(2\pi) \approx 3.2 \, \textrm{MHz}$. Optical microscope images of the sample are shown in Fig.~\ref{fig1}(a).

\begin{figure}[t!]
\includegraphics[width=1 \columnwidth]{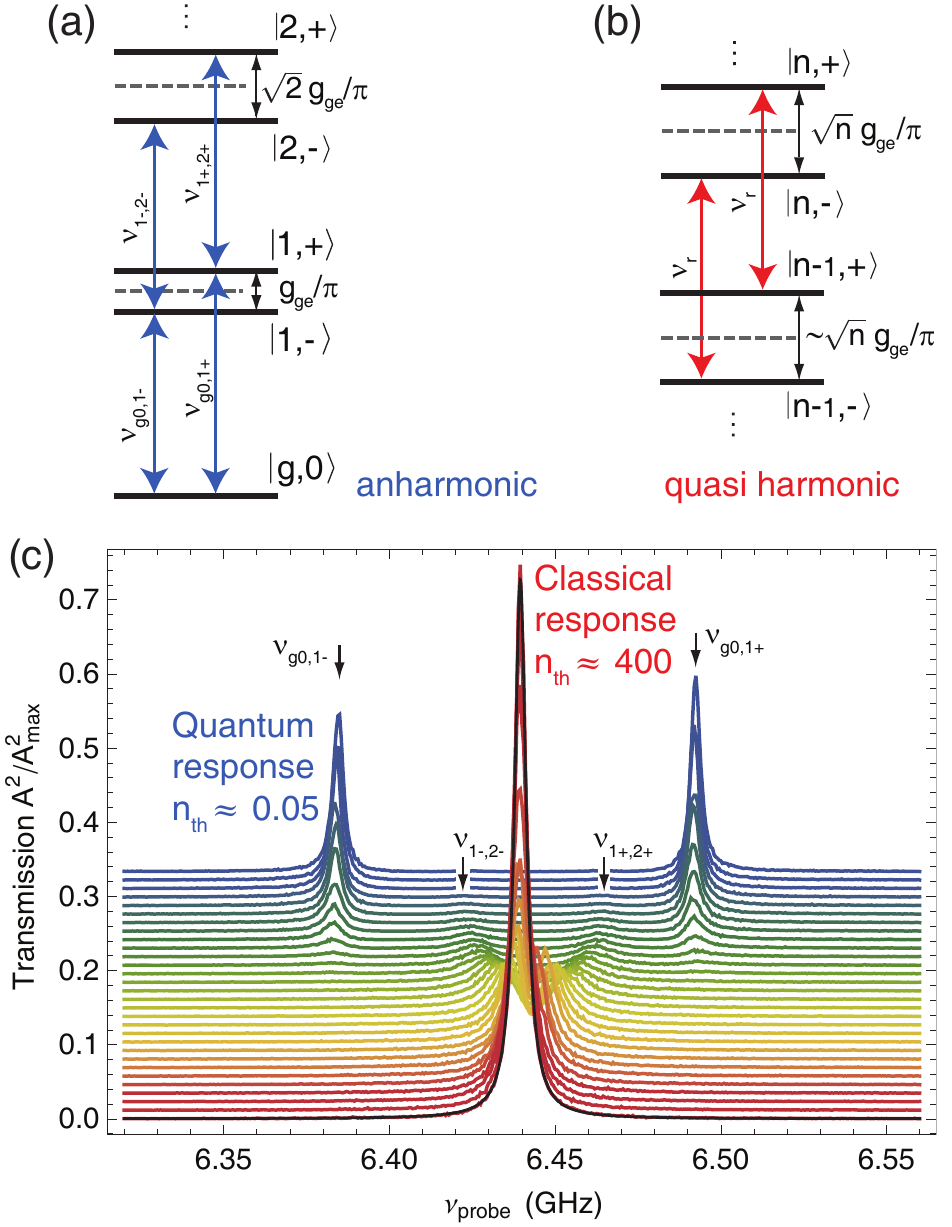}
\caption{(a), Energy level diagram of dipole coupled dressed states $|n,\pm\rangle$ for excitation numbers $n=$ 0, 1 and 2 (black lines), uncoupled qubit and cavity energies (grey dashed lines) and allowed transitions at frequencies $\nu_{\textrm{g}0,1\pm}$ and $\nu_{1\pm,2\pm}$ (blue arrows). (b), Dressed state diagram for large excitation numbers $n > 280$ and allowed transitions at the resonator frequency $\nu_\textrm{r}$ (red arrows). (c), Measured cavity transmission $A^2/A^2_\textrm{max}$ for intracavity thermal photon numbers $0.05$ (blue) $\lesssim n_\textrm{th} \lesssim 400$ (red) and a fit to a Lorentzian line (black). Measured data sets are normalized to the transmission amplitude $A_{\rm{max}}$ obtained when the qubit is maximally detuned. The data is offset and colored for better visibility, see inset in Fig.~\ref{fig5} for color code.} \label{fig2}
\end{figure}

In our experimental setup [Fig.~\ref{fig1}(b)] the coupled qubit/cavity system is prepared in its ground state $|g,0\rangle$ [Fig.~\ref{fig2}(a)], with close to $n=0$ photons in the resonator by cooling the sample to below $20\, \textrm{mK}$ in a dilution refrigerator. Instead of increasing the physical temperature of the sample to control the thermal occupation of the cavity, we apply thermal radiation only at its input. We approximate the one-dimensional Planck spectrum of the thermal field as constant with a power spectral density $S_\textrm{n}$ in the small relevant bandwidth around $\nu_r$. $S_\textrm{n}$ is controlled by applying broadband microwave frequency white noise of controlled amplitude. In this case, we can assign an intra-cavity thermal photon number $n_\textrm{th}=[\exp{(\hbar \omega_\textrm{r} / k_\textrm{B} T_\textrm{c})}-1]^{-1}$ and thus an equivalent cavity field temperature $T_\textrm{c}$ to the externally applied white noise. Owing to the high internal quality factor of the resonator, the field does not thermalize on the chip, which allows us to control the effective temperature of the resonator field to up to $T_\textrm{c} \sim 100 \, \rm{K}$. In addition our setup allows for the phase sensitive detection of the quadrature amplitudes of a weak coherent probe tone that populates the resonator with $n_{\rm{probe}} \lesssim 0.1$ photons on average while effectively rejecting all uncorrelated thermal radiation from the detection system.

At the lowest measured cavity field temperature $T_c \sim 100 \, \rm{mK}$, a clear vacuum Rabi mode splitting is observed in a linear response cavity transmission measurement, see Fig.~\ref{fig2}(c) (dark blue lines). Two Lorentzian lines characteristic for the dressed states $|n,\pm\rangle= \left(|g,n\rangle \pm |e,n-1\rangle\right)/\sqrt{2}$ with excitation number $n=1$ separated by twice the dipole coupling strength $g_\textrm{g,e}/(2\pi)= 54\, \textrm{MHz}$ are clearly observed at the frequencies $\nu_{\textrm{g}0,1\pm}$. This indicates that the strong coupling regime, $g_\textrm{g,e} \gg \kappa$, $\gamma$, with the qubit relaxation rate $\gamma/(2\pi) \approx 0.6\, \textrm{MHz}$, is realized.

\begin{figure*}[t!]
\includegraphics[width=1.90 \columnwidth]{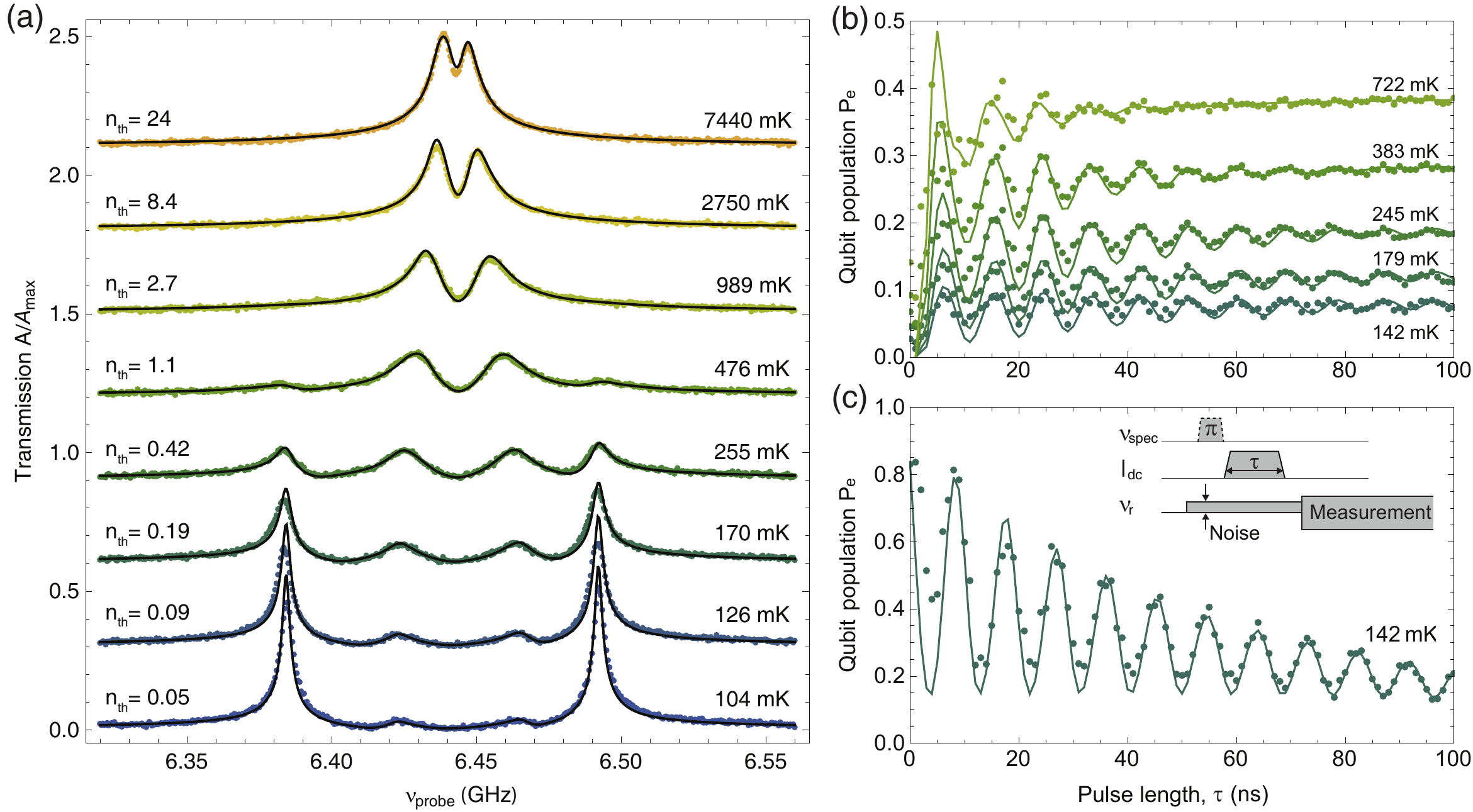}
\caption{(a) Measured (dots) and calculated (lines) cavity transmission (same data as in Fig.~2(c)) shown for applied $S_\textrm{n}=-221.5$~ \textrm{dBm}/\textrm{Hz} (bottom) to $-190$~\textrm{dBm}/\textrm{Hz} (top) in steps of 4.5~\textrm{dBm}/\textrm{Hz}. Extracted thermal photon numbers $n_\textrm{th}$ and cavity temperatures $T_\textrm{c}$ are indicated. (b) Measured qubit excited state population $P_\textrm{e}$ as a function of the resonant cavity interaction time $\tau$ (dots) and master equation simulation (lines) for $S_\textrm{n}=-214$~ \textrm{dBm}/\textrm{Hz} to $-202$~\textrm{dBm}/\textrm{Hz} in steps of 3~\textrm{dBm}/\textrm{Hz}. (c), Similar measurement as in (b) with qubit prepared in the excited state for $S_\textrm{n}=-214$~\textrm{dBm}/\textrm{Hz}. Inset: Vacuum Rabi pulse sequence.} \label{fig3}
\end{figure*}

While the observation of the simple vacuum Rabi mode splitting for $n=1$ could in principle be interpreted as a normal mode splitting using a semiclassical model, the observation of additional transitions to higher excited dressed states with coupling increased by a factor $\sqrt{n}$ \cite{Schuster2008,Fink2008,Bishop2009a,Fink2009b} is an unambiguous signature for the quantum nature of the system \cite{Agarwal1986,Agarwal1991,Tian1992}. In our experiment these transitions are observed at frequencies $\nu_{1\pm,2\pm}$, see the level diagram in Fig.~\ref{fig2}(a) and measurements in Fig.~\ref{fig2}(c) (green lines) [and Fig.~\ref{fig3}(a) (blue and green lines)], when slightly increasing the effective cavity temperature and thus populating excited states of the Jaynes-Cummings ladder. In this case the weak coherent probe tone excites transitions between higher levels.

At high temperatures the Jaynes-Cummings system is excited to high quantum numbers and many transitions can be accessed by the weak probe tone \cite{Agarwal1986,Agarwal1991,Cirac1991,Tian1992,Rau2004a}. The resulting linear response transmission spectrum can be understood as a sum of individual level to level transitions that overlap. At a mean thermal photon number of $n_\textrm{th} > 1$ the transitions from the ground state $\nu_{\textrm{g}0,1\pm}$ are almost saturated and therefore only weakly contribute to the observed cavity transmission spectrum (Fig.~\ref{fig2}(c), yellow lines). At even larger effective temperatures all available transition frequencies are densely spaced close to the resonator frequency, see level diagram in Fig.~\ref{fig2}(b) and measurement in Fig.~\ref{fig2}(c) (red lines). At these temperatures all transitions which would show the nonlinear $\sqrt{n}$ signature of the single qubit--photon interaction are saturated and the transmission spectrum does not carry any information about the intrinsic quantum nature of the system.

The resulting harmonic spectrum resembles that of two uncoupled linear oscillators, a situation where not the absolute energies but the energy differences of the system are degenerate \cite{Maassen2002,Rau2004a}. In contrast to the regime of strong coherent driving, where nonclassicality can persist at large excitation numbers \cite{Alsing1991,Armen2009}, the phase noise of the thermal field used in our experiments renders the two high-excitation dressed-state ladders indistinguishable. As a result at high temperatures the spectrum is that of a classical resonator and we find excellent agreement with a Lorentzian line fit of width $\kappa_{\textrm{eff}}/(2\pi)=4\, \textrm{MHz}$ close to the intrinsic resonator line width (Fig.~\ref{fig2}(c), black line).

A quantitative understanding of the measured results is obtained by numerically solving the Markovian master equation
\begin{equation}\label{master}
\begin{split}
\mathcal{L}[\hat\rho]&=
-\frac{i}{\hbar}[\hat{\mathcal{H}},\hat\rho]
+ (n_\textrm{th}+1)\kappa\mathcal{D}[\hat{a}]\hat\rho\\
&+ \gamma\ \mathcal{D} \left [\sum_{l=\textrm{e,f,...}} \frac{g_{l-1,l}}{g_{g,e}}\ \hat{\sigma}_{l-1,l} \right ]\hat\rho + n_\textrm{th}\kappa\mathcal{D}[\hat{a}^\dagger]\hat\rho \,,
\end{split}
\end{equation}
where the thermal photon number $n_\textrm{th}$ or equivalently the cavity field temperature $T_\textrm{c}$ is extracted as the only fit parameter. Here the coherent dynamics is described by the Jaynes-Cummings Hamiltonian $\hat{\mathcal{H}}$ with $l$ transmon levels, see \cite{Koch2007}, and without a drive term, which is justified in the linear response limit. The three damping terms model the loss of cavity photons at rate $\kappa$, the intrinsic relaxation of the transmon excited state $|e\rangle$ at rate $\gamma$ and the creation of cavity photons due to a thermal bath with occupation $n_\textrm{th}$. $\mathcal{D}$ is the usual Lindblad damping superoperator. Note that neither qubit dephasing nor an independent qubit thermal bath has significant influence on the agreement between the theory and the presented spectroscopic results. Therefore, these two terms have been omitted. Equation~(\ref{master}) is solved by exact diagonalization and the transmission spectrum is calculated similar to Ref.~\cite{Rau2004a}.

\begin{figure}[t!]
\includegraphics[width=1.0 \columnwidth]{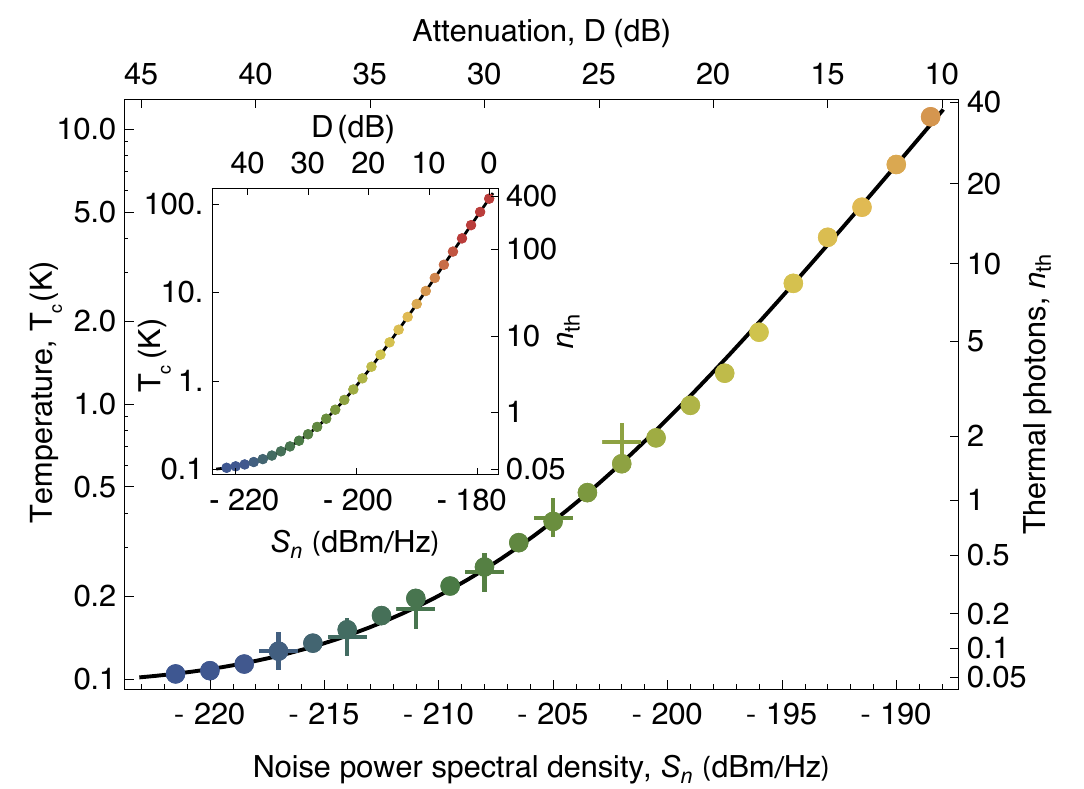}
\caption{Temperature $T_\textrm{c}$ and photon number $n_{\rm{th}}$ extracted from measured Rabi spectra (full dots) and Rabi oscillations (crosses) versus noise power spectral density $S_\textrm{n}$ or equivalently attenuation $D$.  Theory (black line). Inset: extrapolation to higher $T_\textrm{c}$ provides  temperature scale (color coded) for the measurements shown in Figs.~\ref{fig2}(c) and \ref{fig3}.} \label{fig5}
\end{figure}

We observe that the measured and calculated spectra agree very well over a large range of applied thermal noise powers (Fig.~\ref{fig3}). This allows to extract the effective cavity field temperature (Fig.~\ref{fig5}). For mean thermal photon numbers $n_\textrm{th}\gtrsim30$ the accuracy of our numerical calculations is limited by the finite size of the Hilbert space spanned by 200 resonator states and 6 transmon states. In the large photon number limit the qubit only negligibly perturbs the empty cavity Lorentzian spectrum [\figref{fig2}(c)], because the cavity and cavity dissipation terms in Eq.~\eqref{master} scale linearly with $n_\textrm{}$, while the qubit photon coupling term scales only with $\sqrt{n_\textrm{}}$, see also \cite{Savage1989}. In the limit where the coupling is smaller than the damping $\sqrt{n} g_\textrm{g,e} < n \kappa$ the dominant terms in Eq.~\eqref{master} describe a damped harmonic oscillator with a Lorentzian spectrum. A classical harmonic oscillator is furthermore a good description of our cavity QED system if the relative $\sqrt{n}$ nonlinearity, i.e.~the frequency shift due to the qubit, is smaller than the relevant dissipation rate $2\,g_\textrm{g,e} (\sqrt{n_{}+1}-\sqrt{n_{}}) < \kappa$. For large photon numbers these two criteria are equivalent and in our experiment, the classical limit is entered for mean thermal photon numbers $n_\textrm{th} > (g_\textrm{g,e}/\kappa)^2\approx 280$.

We have also performed time domain vacuum Rabi oscillation measurements in the presence of thermal photons. In these experiments, we first detune the qubit from the resonator by $0.5\,\textrm{GHz}$ where the applied quasi thermal noise is filtered and does not affect the qubit state. We then apply an amplitude shaped current pulse $I_{\rm{dc}}$ via an on-chip flux line, see sample and setup in Fig.~\ref{fig1}, to tune the qubit in resonance with the cavity field for a variable time $\tau$, see pulse sequence in inset of Fig.~\ref{fig3}(c). The qubit excited state population $P_\textrm{e}$ is then determined in a dispersive measurement \cite{Bianchetti2009}.

For the lowest cavity temperatures we observe low contrast, long coherence vacuum Rabi oscillations induced by the weak thermal field only [Fig.~\ref{fig3}(b)]. When the number of thermal photons is increased the amplitude of the coherent oscillations increases while their coherence time decreases, which is expected for a thermal distribution of photon numbers.
At long interaction times $\tau$, the qubit population shows a temperature dependent saturation that approaches the value of $0.5$ when the qubit is strongly driven by the thermal field at high temperatures. We have repeated the experiment preparing the qubit initially in the excited state using a $\pi$-pulse and measuring vacuum Rabi oscillations at low applied noise power [Fig.~\ref{fig3}(c)]. In this case high contrast oscillations are observed albeit with a small offset in the qubit population visible at large $\tau$ caused by the thermal field. From these data the cavity temperature has been extracted by solving a time dependent master equation, see solid lines in Figs.~\ref{fig3}(b) and (c). In this case 2 qubit states and up to 6 cavity states are considered and qubit dephasing 
is included. The deviations at short times $\tau$ are due to imperfections in the qubit tuning pulses.

The cavity field temperatures $T_c$ extracted from both spectroscopic (full dots) and time-resolved measurements (crosses) plotted on a logarithmic scale versus the applied noise spectral density $S_\textrm{n}$ are in close agreement (Fig.~\ref{fig5}). Both data sets are explained consistently considering that the cavity thermal photon number $n_\textrm{th}=S_\textrm{n}/\hbar\omega+n_0 $ is the sum of the applied quasi thermal noise $S_\textrm{n}/\hbar\omega$ expressed as a photon number and a small thermal background field of $n_0=0.04$ photons ($T_0=95~\textrm{mK}$). The measured intracavity photon number $n_\textrm{th}$ is consistent with the noise applied to the input of our setup and the expected attenuation of the line and also with calibration measurements based on the known qubit ac-Stark shift \cite{Schuster2005}. The background photon number is set by incomplete thermalization of the input line and can easily be decreased by adding additional cold attenuation.

The inset of Fig.~\ref{fig5} shows an extrapolation of the temperature to the largest applied noise powers and provides a calibration for the measurements shown in Fig.~\ref{fig2}(c) where the same color code is used (also used in Figs.~\ref{fig3} and \ref{fig5}). Accordingly, the measurements at the highest temperatures were conducted with approximately $\sim 370$ thermal photons in the cavity. This corresponds to an effective cavity temperature of $T_\textrm{c}\sim115~\textrm{K}$. We clearly observe the transition to a classical harmonic resonator at $n_\textrm{th}\approx370>(g_\textrm{g,e}/\kappa)^2\approx 280$
as indicated by the Lorentzian spectrum  [Fig.\ref{fig2}(c)].

We have demonstrated a quantitative understanding of the transition from the quantum to the classical response of a cavity QED system. Moreover, we have presented a viable approach to measure the temperature of an electromagnetic field in a cavity over a wide range.  A different approach to extract the resonator temperature in the dispersive regime of circuit QED has been proposed \cite{Clerk2007} and measurements of the qubit temperature have been demonstrated \cite{Laloy2009}. Related experiments have been performed using a semiconductor cavity QED system \cite{Laucht2009}. In these systems, entanglement and decoherence at elevated temperatures can be studied in future experiments in the context of quantum information.

\begin{acknowledgments}
We thank S.~Michels for his contribution to the project, J.~Koch and A.~Shnirman for discussions and G.~S.~Agarwal and H.~J.~Carmichael for comments on the manuscript. We acknowledge the group of M.~Siegel at the University of Karlsruhe for the preparation of Niobium films. This work was supported by SNF, EuroSQIP and ETHZ. L.~S.~B.~acknowledges support by LPS/NSA under ARO contract No.~W911NF-05-1-0365 and the NSF under grant No.~DMR-0653377.
\end{acknowledgments}

\end{document}